\documentclass[twocolumn,aps,prd,nofootinbib]{revtex4-1}
\usepackage[colorlinks=true,linkcolor=black,citecolor=blue,urlcolor=black]{hyperref}
\usepackage{graphicx,soul}
\usepackage{subfigure}
\usepackage{natbib}
\usepackage{epsfig}
\usepackage{amstext,amsmath,amsfonts,amssymb,amsthm}
\usepackage[format=plain,justification=centerlast]{caption}
\usepackage{color}

\usepackage{verbatim}
\usepackage{lipsum}
\usepackage{color,array,wrapfig}
\usepackage{caption}
\def\be{\begin{equation}}
\def\ee{\end{equation}}
\def\bes{\begin{eqnarray}}
\def\ees{\end{eqnarray}}
\def\ba{\begin{align}}
\def\ea{\end{align}}
\def\bwt{\begin{widetext}}
\def\ewt{\end{widetext}}

\def\aa{\alpha}

\newbox\one
\newbox\two
\long\def\loremlines#1{%
    \setbox\one=\vbox {%
       Test.\footnote{a footnote}%
      \lipsum\footnote{Another footnote.}%
     }
   \setbox\two=\vsplit\one to #1\baselineskip
   \unvbox\two}

\begin{document}
\title{Challenges in Inflationary Magnetogenesis: Constraints from Strong Coupling, Backreaction and the Schwinger Effect}

\author{Ramkishor Sharma$^{1}$}
\email{rsharma@physics.du.ac.in, sharmaram.du@gmail.com}
\author{Sandhya Jagannathan$^{1}$} 
\email{sandhya0892@gmail.com} 
\author{T. R.Seshadri$^{1}$}
\email{trs@physics.du.ac.in}
\author{Kandaswamy Subramanian$^{2}$} 
\email{kandu@iucaa.in} 
\affiliation{$^{1}$Department of Physics \& Astrophysics, University of Delhi, New Delhi$-$110007 India.}
\affiliation{$^{2}$ IUCAA, Post Bag 4, Pune University Campus, Ganeshkhind, Pune$-$411007 India.}

\vskip 1cm
\begin{abstract}
Models of inflationary magnetogenesis 
with a coupling to the electromagnetic action of the form
$f^2 F_{\mu\nu}F^{\mu\nu}$, are known to suffer from several problems.
These include the strong coupling problem, the back reaction problem 
and also strong constraints due to Schwinger effect. 
We propose a model which resolves all these issues. In our model, 
the coupling function, $f$, grows during inflation and transits to 
a decaying phase
 post inflation.
This evolutionary behaviour is chosen so as to avoid the problem of 
strong coupling. By assuming a suitable power law form of the coupling 
function, we can also neglect back reaction effects during inflation. 
To avoid back reaction post-inflation, we find that the reheating temperature 
is restricted to be below $ \approx 1.7 \times 10^4$ GeV. 
The magnetic energy spectrum is predicted to be non-helical and generically
blue. The estimated present day magnetic field strength and the corresponding
coherence length taking reheating at the QCD epoch(150 MeV) are 
$ 1.4 \times 10^{-12}$ G and  $6.1 \times 10^{-4}$ Mpc, respectively. 
This is obtained after taking account of
nonlinear processing over and above the flux freezing evolution after reheating.
If we consider also the possibility of a non-helical inverse transfer, 
as indicated in direct numerical simulations, the coherence length 
and the magnetic field strength are even larger.
In all cases mentioned above, 
the magnetic fields generated in our models 
satisfy the $\gamma$-ray bound below a certain reheating temperature.
\end{abstract}

\maketitle
\section{Introduction}
Magnetic fields have been observed over a wide range of length scales in the universe. They have been detected in galaxies, galaxy clusters and even in intergalactic voids \cite{r-beck,clarke,widrow,neronov}. Gamma ray observations have put a lower bound on the intergalactic magnetic fields of the order $ 10^{-15}$G for fields coherent on Mpc scales \cite{neronov}. The mechanism responsible for the origin of these fields is not yet clearly understood although considerable work has been done in this regard. There are two approaches in this context. One approach attributes the origin of seed fields to astrophysical batteries which are then further amplified by flux freezing and dynamo action \cite{zel, Shukurov,bks,kulsurd2008}. The other approach proposes a primordial origin of seed fields. The proponents of this scenario suggest that the generation of seed fields can possibly occur due to processes in early universe 
during inflation \cite{turner-widrow, ratra, martin-yokoyama, kobayashi2014, atmjeet2014, Campanelli2015, 1475-7516-2015-03-040}, electroweak \cite{vachaspati, Sigl:1996dm} or QCD phase transitions \cite{kisslinger,qcd} (for reviews, see  \cite{grasso, dner, kandu2010, 2011PhR...505....1K, kandu2016}). Primordial origin would be more favoured if the observations of magnetic fields in voids is firmed up.

Inflation offers a natural setting to explain observed large scale magnetic fields since large scale features
 emerge naturally from the theory. However, the magnetic field strength generated during inflation decreases with the expansion factor $a(t)$, very rapidly as $ B \propto 1/{a^{2}}$, as the standard EM action is conformally invariant. This results in a very low present day strength which is far below the value required to even seed the dynamo. Hence conformal invariance needs to be broken to obtain fields of sufficient strengths that are observed today. Inflation provides many scenarios where this is possible, one of them is in which the inflaton field couples to the kinetic term of the electromagnetic (EM) field \cite{ratra}.
However this model is also beset with several problems such as the strong coupling and the back reaction problem. Strong coupling problem occurs if the effective electric charge is high during some epoch of inflation thereby making the perturbative calculations of the EM test field untrustworthy. We will address this issue in more detail in Section \ref{ascp}. In some cases, during inflation, the electric and magnetic energy density can overshoot the background energy density. This can end inflation as well as suppress the production of magnetic fields. This is known as the back reaction problem. Models of low energy scale inflation have been suggested where the back reaction problem has been tackled \cite{rajeev}. In some of these models, however, Schwinger effect constraint poses a problem \cite{kobayashi:2014}. Schwinger mechanism is the production of charged particles due to electric fields.
If the electric field is high enough during inflation, it can generate charged particles. This can lead to the conductivity becoming very high. This will affect the EM field and can result in a low strength of the magnetic field today.

In this paper, we propose a model to tackle all the above problems besieging inflationary magnetogenesis. In our model, we allow the coupling function (function of the inflaton which breaks conformal invariance, $f$) to evolve during inflation and also during a matter dominated era from the end of inflation to reheating.
By demanding that EM field energy density does not back react on the background even after inflation, we get a constraint on the scale of inflation and the reheating temperature. We give several scenarios which satisfy all the three constraints. 
They all imply a low reheating scale and a blue spectra for the generated fields, with a sub-horizon coherence length. Therefore one has to consider the nonlinear effects discussed by Banerjee and Jedamzik \cite{jedamzik} during the radiation dominated era after reheating. The field strength decays while the coherence length increases due to this evolution.
The generated field strengths and coherence scales in several of our models are 
consistent with the potential lower limits from $\gamma$-ray observations.

The outline of the paper is as follows: In Section \ref{emdi} and \ref{maed}, we provide a general background about the evolution of EM fields during inflation and further state the resultant form of power spectra of magnetic and electric energy density. In Section \ref{se}, we discuss the possible constraints on magnetogenesis arising out of Schwinger effect and show how these can be satisfied. In Section \ref{ascp}, we address the issue of strong coupling and give a detailed account of the model which solves this problem. Section \ref{result} gives the predictions arising out of our model, without taking nonlinear effects into consideration. In Section \ref{nl}, we add nonlinear effects into our model, also discussing the mechanism of inverse transfer briefly in this context.
Further in Section \ref{discussion}, we discuss whether our results conform with the constraints obtained from gamma ray observations. Our conclusions are given in Section \ref{conclusion}.
\section{Evolution of Electromagnetic Field during inflation}\label{emdi}
The standard Maxwell action is invariant under conformal transformation \cite{parker} and FRW metric is conformally flat. Due to this the electromagnetic (EM) fluctuations decay rapidly as the square of the scale factor. Hence, breaking of conformal invariance is necessary for inflationary magnetogenesis \cite{turner-widrow}. We start with the action for the EM field in which the conformal invariance is explicitly broken by introducing a time dependent function $f^2(\phi)$, where $\phi$ is the inflaton field, coupled to the kinetic term ($F^{\mu \nu}F_{\mu \nu}$) in the action \cite{ratra}.
\begin{align}
S &=-\int\sqrt{-g} d^4x [f^2(\phi)\frac{1}{16 \pi}{F_{\mu\nu}}{F^{\mu\nu}} + j^{\mu}A_{\mu}]\nonumber\\ 
 &-\int\sqrt{-g} d^4x \Big[\frac{1}{2}\partial^\nu \phi \partial_\nu \phi + V(\phi)\Big]\label{actionfull}.
\end{align}
Here $F_{\mu\nu}=\partial_\mu A_\nu - \partial_\nu A_\mu$, where $A_\mu$ is the EM 4-potential. The term $j^\mu A_\mu$ represents the interaction where $j^\mu$ is the four current density. The second term in the action incorporates the evolution of the inflaton field. In this paper we have adopted Greek indices $\mu, \nu....$ to represent space-time coordinates and Roman indices $i,j,k....$ to represent spatial coordinates. We follow the metric convention $g_{\mu\nu}=diag(-,+,+,+)$. 
To begin with, we neglect the interaction term and assume that there are no free charges. 
Varying the action with respect to the EM 4-potential, we obtain the following modified form of Maxwell's equations.
\begin{equation}
[f^2 F^{\mu\nu}]_{;\nu}=\frac{1}{\sqrt{-g}}\frac{\partial}{\partial x^\nu}[\sqrt{-g}g^{\mu\alpha}g^{\nu\beta} f^2(\phi) F_{\alpha\beta}]=0 \label{ME}.
\end{equation}
Varying the action with respect to the scalar field we obtain the following equation.
\begin{equation}
\frac{1}{\sqrt{-g}} \frac{\partial}{\partial x^\nu}\Big[\sqrt{-g} g^{\mu\nu}\partial_\mu \phi \Big] -\frac{dV}{d\phi}=\frac{f}{8 \pi}\frac{df}{d\phi} F_{\mu\nu} F^{\mu\nu}.
\end{equation}
Here EM field is assumed to be a test field and hence, it will not affect the evolution of the background which is dominated by the scalar field potential during inflation. The scalar field $\phi$ is assumed to be homogeneous, having only time dependence. Adhering to the homogeneity and isotropy of the universe, we work in FRW space-time and further assume it to be spatially flat,
\begin{eqnarray}
ds^2&=&-dt^2+a^2(t)[dx^2+dy^2+dz^2]\nonumber\\
&=& a^2 (\eta)[-d\eta^2+dx^2+dy^2+dz^2].
\end{eqnarray}
In this new coordinate system $(\eta,x,y,z)$, $\eta $ denotes conformal time. Further, to solve Eq.(\ref{ME}), it is convenient to adopt the Coulomb gauge,
$$\partial_j A^j =0 \quad \quad A_0=0.$$
We can express Eq.(\ref{ME}) for $\mu=i$ as,
\begin{equation}\label{Aeq}
A_i''+2 \frac{f'}{f} A_i'-a^2 \partial_j \partial^j A_i=0.
\end{equation}
Here prime ($'$) denotes derivative with respect 
to $\eta$ and $\partial^j$ is defined as $\partial^j\equiv g^{jk}\partial_k=a^{-2 }\eta^{jk} \partial_k$. Promoting $A_i$ to an operator and imposing the quantization condition, we expand $A_i$ in terms of creation and annihilation operators in Fourier space 
\cite{martin-yokoyama,kandu2010}. 
The evolution equation for the 
corresponding 
mode function 
in Fourier space $A(k,\eta)$, 
can be obtained  
from 
Eq.(\ref{Aeq}), and in terms of a new variable $\bar A \equiv a A(k,\eta)$, becomes 
\begin{equation}
\bar A''+2 \frac{f'}{f} \bar A'+ k^2\bar A=0. \label{genA}
\end{equation}
Here $k$ is the comoving wave number. We can re-express the above equation in the form of a harmonic oscillator equation with a time dependent frequency. To do this, we further define a new variable $\mathcal A \equiv f\bar{A}(k,\eta)$. The equation of motion in terms of this new variable is,
\begin{equation}
\mathcal A''(k,\eta)+\Big(k^2-\frac{f''}{f} \Big) \mathcal A(k,\eta)=0.   \label{scriptAeq}
\end{equation}
Before we solve the above equation for a particular $f(\phi)$, we first define the magnetic and electric energy density respectively as,
\begin{align}
\rho_B = \langle0|T^{B}_{\mu \nu}u^{\mu}u^{\nu}|0\rangle \quad \text{and} \quad \rho_E = \langle0|T^{E}_{\mu \nu}u^{\mu}u^{\nu}|0\rangle ,\label{rhobae}
\end{align}
where  $ \rho_B $ and $\rho_E$ are defined as the vacuum expectation value of the respective energy momentum tensors $T^{B}_{\mu \nu}$ and $T^{E}_{\mu \nu}$, measured by the fundamental observers. The velocity of these observers $u^\mu$ is specified as $(1/a,0,0,0)$.
In our analysis, we work with the spectral energy densities of magnetic and electric fields, 
$(d\rho_B(k,\eta)/d\ln k)$ and $(d\rho_E(k,\eta)/d\ln k)$.
These spectral energy densities can be obtained from the Fourier transform of $\rho_B$ and $\rho_E$ in Eq.(\ref{rhobae}) 
\cite{martin-yokoyama,kandu2010},
\begin{eqnarray}
\frac{d\rho_B(k,\eta)}{d \ln k} &=&\frac{1}{2 \pi^2} \frac{k^5}{a^4} |\mathcal A(k,\eta)|^2 \label{rhob}\\
\frac{d\rho_E(k,\eta)}{d \ln k} &=&\frac{f^2}{2 \pi^2} \frac{k^3}{a^4} \Big|\Big[\frac{\mathcal A(k,\eta)}{f}\Big]'\Big|^2. \label{rhoe}
\end{eqnarray}
\section{Magnetic and Electric Energy Density during Inflation}\label{maed}
In this analysis, we assume the background to be de Sitter during inflation. The evolution of the scale factor $a(\eta)$ with conformal time $\eta$ during de Sitter, is given by,
\begin{equation}
a = \frac{-1}{H_{f}\eta},
\end{equation}
where $H_{f}=a'/a^2$ is the Hubble parameter during inflation.
In our normalization, $a \equiv a_{i} = 1$ at $\eta_{i} = -1/H_{f}$, where $\eta_{i}$ and $a_{i}$ are the conformal time and the scale factor at the beginning of inflation respectively. 
To solve Eq.(\ref{scriptAeq}), 
we need to know how $f(\phi)$ evolves with time or with expansion factor. 
We assume $f$ evolves as, 
\begin{equation}
f(a) = f_i\left(\frac{a}{a_i}\right)^{{\aa}}.
\end{equation}
For any model of inflation this form of $f$ can be chosen by adopting appropriate functions of $\phi$ \cite{martin-yokoyama}. 
For the above mentioned form of $f(\eta)$, Eq.(\ref{scriptAeq}) reduces to,
\begin{equation}
\mathcal A''(k,\eta)+\Big(k^2- \frac{\aa(\aa+1)}{\eta^2} \Big) \mathcal A(k,\eta)=0.   \label{scriptA}
\end{equation}
The general solution of this equation is given by,
\begin{equation}
\mathcal A_1=\sqrt{-k\eta} \big[c_1(k)J_{-\aa-\frac{1}{2}}(-k\eta)+c_2 (k) J_{\aa+\frac{1}{2}}(-k\eta)\big]. \label{A}
\end{equation}
The constants $c_1$ and $c_2$ are determined by matching the solution above with the mode functions corresponding to the Bunch-Davies vacuum in the limit of $(- k \eta) \rightarrow \infty $,
\begin{align}
c_1&=\sqrt{\frac{\pi}{4k}}\frac{\exp\frac{i\pi \aa}{2}}{\cos(-\pi \aa)}, \quad c_2=\sqrt{\frac{\pi}{4k}} \frac{\exp\frac{i\pi(-\aa+1)}{2}}{\cos(-\pi\aa)}. 
\end{align}
Using Eq.(\ref{rhob}) and Eq.(\ref{A}), the magnetic energy density spectrum in the super-horizon limit $(-k\eta)<<1$ becomes,
\begin{eqnarray}
\frac{d\rho_B}{d \ln k}&\approx&\frac{\mathcal{F}(n)}{2 \pi^2} H_f^4 (-k\eta)^{4+2n}\nonumber\\
&\approx &\frac{\mathcal{F}(n)}{2 \pi^2} H_f^4 \left(\frac{k}{aH_{f}}\right)^{4+2n}  \label{rhob2}
\end{eqnarray}
where,
\begin{equation}
n=-\aa~\text{if}~\aa \geq -1/2~\text{and}~n=1+\aa~\text{if}~\aa \leq -1/2 \nonumber  \label{n}
\end{equation}
and
\begin{equation}
\mathcal{F}(n)=\frac{\pi}{2^{2n+1}\Gamma^2(n+1/2)\cos^2(\pi n)}.  \label{F}
\end{equation}
Similarly we can determine the spectral electric energy density as,
\begin{equation}
\frac{d\rho_E}{d \ln k}\approx \frac{\mathcal{G}(m)}{2 \pi^2} H_f^4 (-k\eta)^{4+2m}. \label{rhoea}
\end{equation}
Here, $\cal G$ is given by, 
\begin{equation}
\mathcal{G}(m)=\frac{\pi}{2^{2m+3}\Gamma^2(m+3/2)\cos^2(\pi m)}  \label{9}
\end{equation}
and $m$ takes the values,
\begin{equation}
m=-\aa+1~\text{if}~\aa\geq 1/2~\text{and} ~m=\aa~ \text{if}~\aa \leq 1/2. \label{mval}
\end{equation}
First we explore the possibility of scale invariant magnetic field spectrum. However, we also note that scale invariance for magnetic field spectrum does not imply the scale invariance of electric field spectrum. There are two possible values of $\aa$ for scale invariant magnetic field spectrum, namely, $\aa=-3$ and $\aa=2$ .
 
For the first branch, $\aa=-3$, we have from Eq.(\ref{mval}) $4+2m=-2$. In this case, as $(-k\eta)\rightarrow 0$ (i.e. towards the end of inflation), the electric density increases rapidly as $\rho_E \propto (-k\eta)^{-2}\rightarrow \infty$. In this case the model runs into difficulties as the electric energy density would eventually exceed the inflaton energy density in the universe even before sufficient inflation.
 This problem is known as the back reaction problem. Therefore $\aa=-3$ 
branch 
for the generation of magnetic fields
is strongly constrained. 
Also production of large electric fields can give rise to finite conductivity due to  Schwinger effect during inflation which can further stop the generation of magnetic fields as discussed in \cite{kobayashi:2014}.

This motivates us to rather choose the  
second branch, $\aa=2$, 
for which
we have from Eq.(\ref{mval}) $4+2m=2$. As $(-k\eta)\rightarrow 0$, $\rho_E \propto (-k\eta)^2\rightarrow 0$. This branch does not suffer from any back reaction effects, although strong coupling 
could potentially pose
a problem here as discussed in Section \ref{ascp}. In Section \ref{ascp}, we also suggest a model in which strong coupling problem is avoided. Also the issue of finite conductivity arising from  particle production due to Schwinger mechanism does not pose a problem for this branch. This is shown explicitly in the following section. 
\section{Schwinger effect constraint on Magnetogenesis}\label{se}
It is well known that electric fields can produce charged particles out of the vacuum due to Schwinger mechanism \cite{schwinger, sharma2017}. These particles can give rise to a finite conductivity depending on the electric field strength,
resulting in the electric field being damped and the growth of magnetic fields being frozen \cite{kobayashi:2014}.  
Kobayashi and Afshordi 
\cite{kobayashi:2014}
have determined the conductivity due to charged particles produced via Schwinger mechanism in de Sitter space-time for different strengths of electric fields. Then using the value of conductivity they have put constraints on inflationary magnetogenesis models. In our analysis we had neglected the interaction term but to check the effect of conductivity we need to reinstate the interaction term back into the equation of motion.

After including the interaction term in Eq.\eqref{Aeq} we get,
\begin{equation}
A_i''+\Big(2 \frac{f'}{f}+\frac{4 \pi a \sigma}{f^2}\Big) A_i'- \partial_j \partial_j A_i=0. \label{7}
\end{equation}
In writing this equation we have taken $j_i=\sigma E_i=-\sigma (1/a)(\partial A_i/\partial \eta)$ where $\sigma$ denotes conductivity due to the charged particles produced. Clearly this would reduce to the previous case (i.e. without interaction) in the limit,
\begin{equation}
\Big|2 \frac{f'}{f}\Big| \gg \Big|\frac{4 \pi a \sigma}{f^2}|. \label{limit}
\end{equation}
We now proceed to evaluate and examine the validity of this inequality 
for the growing branch, $\aa\ge1/2$. 
In this case Eq.(\ref{limit}) translates to
\begin{equation}
\frac{2\sigma \pi}{\aa H_f f^2} \ll  1 \label{8},
\end{equation}
for the Schwinger effect to be unimportant.
To check this bound we need to estimate $\sigma/H_f$, 
which in turn has been shown to depend on the electric field strength \cite{kobayashi:2014}.
For the two cases $|e_N E|\ll H^2$ and $|e_N E|\gg H^2$, we adopt the results derived in \cite{kobayashi:2014}. We have defined $e_{N}$ as the effective charge, $e_{N} = e/f^{2}$. In the limit $f \rightarrow 1$, we get back the standard electric charge $e$.
For $|e_N E|\gg H_f^2$
\begin{equation}
 \frac{\sigma}{H_f} \simeq sgn(E) \frac{1}{12 \pi^2}\frac{|e_N|^3 E}{H_f^2}e^{\frac{\pi m^2}{|e_NE|}}\label{9b}
\end{equation}
For $|e_N E|\ll H_f^2$, the results depend on the mass of the charged scalar field $m$.
If $m > H_f$, we have,
 \begin{equation}
\frac{\sigma}{H_f} \simeq \frac{7}{72 \pi^2}\frac{e_N^2 H_f^2}{m^2} = 
\frac{7}{72 \pi^2}\frac{e^2~ H_f^2}{f^4~ m^2} , \label{10}
 \end{equation}
and if $m < H_f$, we have
\begin{equation}
 \frac{\sigma}{H_f} \simeq \frac{3}{4 \pi^2}\frac{e_N^2 H_f^2}{m^2}
= \frac{3}{4 \pi^2}\frac{e^2~ H_f^2}{f^4~ m^2}. 
\label{11}
\end{equation}
The relations discussed above show how the conductivity behaves in the presence of electric field. It is worth emphasising that 
Kobayashi and Afshordi \cite{kobayashi:2014}, 
have assumed that the charged scalar field and the EM field do not affect the background.

To check which one of the above cases is relevant to our analysis, we need to estimate the electric field strength in our case. Using Eq.(\ref{rhoea}), 
\begin{align}
E^2(a)\equiv 8 \pi \rho_E=\frac{8 \pi}{f^2}\int d \ln k \frac{\mathcal{G}(-\aa+1)}{2 \pi^2} H_f^4 \left(\frac{k}{aH_f}\right)^{6-2\aa}.\label{bb}
\end{align} 
From the above expression we can see that the electric energy density integral diverges for large values of $k$. We are however interested in the length scales that exit the Hubble radius during the inflationary period. This gives a range for the relevant $k$ over which the integral needs to be performed. The largest length scale of interest is the one which exits the horizon at the beginning of inflation. Hence the physical size of this scale should be equal to the Hubble radius during inflation (The Hubble radius is constant during inflation). Hence, comoving $L_{upper}=1/H_f a_i$ which gives $k_{lower} =H_f a_i$. 
At a time $t$ during inflation, the scale factor $a>a_{i}$ and the comoving length exiting the horizon at that time is $L=(H_{f}a)^{-1}$ implying $k_{upper} = H_{f}a$. The smallest length scale of interest is the one which left the horizon at the end of inflation.
Evaluating the integral at any time t during inflation we have,
$L_{lower}=H_f^{-1}/a$ implying $k_{upper}=H_f a$. Evaluating the integral using the above limits, we have,
\begin{eqnarray}
E^2(a)&\approx&\frac{8 \pi}{f^2} \frac{\mathcal{G}(-\aa+1)}{2 \pi^2(6-2\aa)} H_f^4 \left(1-\left(\frac{a_i}{a}\right)^{6-2\aa}\right). \label{6}
\end{eqnarray}
This implies,
\begin{align}
\frac{|e_N E|}{H_f^2}\approx \frac{2 |e|}{f^3} \left(\frac{\mathcal{G}(-\aa+1)}{\pi(6-2\aa)}\right)^{\frac{1}{2}}\left(1-\left(\frac{a_i}{a}\right)^{6-2\aa}\right)^{\frac{1}{2}}.\label{21}
\end{align}
Since $a>a_i$ we can neglect $a_i/{a}$ in the above expression for $1/2 \le \aa<3$. 
For $\aa=3$, Eq.(\ref{21}) modifies to $|e_N E|/H_f^2=(2|e|/f^3) (\mathcal{G}(-2) \ln(a/a_i)/\pi)^{1/2}$.
Further assuming that $f=1$ at the beginning of inflation and that it grows during inflation, we can infer that $|e_N E| \ll H_f^2$ is valid throughout the inflationary regime for $1/2\le\aa\le3$.
As we saw for the case $|e_N E| \ll H_f^2$, there are two possibilities: (i) $m>H_f$ and (ii) $m<H_f$. For $m>H_f$, we have from Eq.(\ref{10}), 
\begin{equation}
  \frac{2\sigma \pi }{\aa H_f f^{2}} \simeq \frac{14}{72 \pi \aa}
\frac{e^2~ H_f^2}{f^6~ m^2}. 
\label{12}
\end{equation}

Since $f>1$ during inflation
and also $m>H_f$ above,
 we can infer from the above expression that $2\sigma \pi /\aa H_f f^{2}$ is always less than 1. Hence, Eq.(\ref{8}) is valid for $m > H_{f} $. This implies that magnetic field generation will not be affected in this case even if we consider the conductivity of the medium.
On similar lines, for the case of $m < H_f$, we have from Eq.(\ref{11}),
\begin{equation}
    \frac{2\sigma \pi}{\aa H_f f^{2}} \simeq \frac{6}{4 \aa \pi}\frac{e^2~ H_f^2}{f^6~ m^2}. \label{13}
\end{equation}
Thus for $m < H_{f}$, even if initially $2 \sigma \pi/ \aa H_{f} f^{2} \gg 1$, as $f \propto a^{\aa}$ grows rapidly in time, one would have $ 2\sigma \pi/ \aa H_{f} f^{2} \ll 1 $ and the effect of Schwinger conductivity would become negligible. 
By keeping in mind the validity of Eq. $\eqref{8}$, we get the following bound for this case: 
\begin{eqnarray}
 H_f>m>\sqrt{\frac{6 e^2}{4 \aa \pi}}\frac{H_f}{f^3}\label{14}
\end{eqnarray}
For $m < H_f$, if $m$ satisfies the above condition, conductivity will not affect inflationary magnetogenesis. This will always be satisfied as $f$ grows much larger than unity.

Thus we can conclude that our selected branch ($1/2\le\aa\le3$) is not affected by the finite conductivity of the charged particles produced due to Schwinger mechanism. The backreacting branch $\aa = -3$ is indeed strongly constrained as already pointed out in \cite{kobayashi:2014}.
\section{Solving the strong coupling problem}\label{ascp}
We mentioned in Section \ref{maed} that $\aa=2$ branch keeps electric field under control. We also saw above that it does not face the strong constraints imposed by the Schwinger effect. However it suffers from 
a variant of
the strong coupling problem. This problem was first pointed out by Demozzi, Mukhanov and Rubinstein \cite{mukhanov2009}. It states that if $f$ grows during inflation and settles down to $f_f=1$ at the end of inflation, the value of $f$ would need to be very small at the beginning of inflation ($f=f_i$). The effective charge defined by, $e_N=e/f_i^2$ would then be very high at the beginning. This would imply that the coupling between charged particles and the EM field would be unacceptably strong. Alternatively, if $f$ decreases from a large value to $f_{f} = 1$, one could in principle avoid this problem. However we already pointed out that 
a scenario where $f$ decreases rapidly, 
suffers from the back reaction problem.

On the other hand, if we assume that the value of $f$ at the beginning of inflation is $f_i=1$, then there will be a large value of $f$ at the end of inflation and hence resulting in a very small value of coupling constant. This as such will not be a problem\cite{kandu2010}. 
However, we would need to ensure that after the end of inflation, the value of $f$ decreases to its pre-inflationary value to restore back the standard couplings. This should happen within a time scale such that it does not affect the known standard physics. If this is achieved, one would have found a way of solving the strong coupling problem.

In our model, we assume that $f$ increases as a power law during de Sitter inflation beginning with a value of unity, after which it decays back to its pre-inflationary value during a matter dominated phase that lasts till reheating. The form of the power law during the inflationary phase is given by, 
\begin{equation}
f=f_1 \propto a^{\aa}~~~~~~~~~~  a_i \le a \le a_f  \label{faalpha}
\end{equation}

and that in the post inflationary epochs by,
\begin{equation}
f=f_2 \propto {a}^{-\beta}~~~~~~~~~ a_f \le a \le a_r .
\end{equation}
Here $ a_f$ and $a_r$ denote the scale factor at the end of inflation and at the end of reheating, respectively. We consider $\aa > 1/2$ in our analysis and keep $ \beta $ as general. Since the value of the function 
$f$ 
is taken to be unity at the beginning of inflation, its value at the end of inflation is $f_f=(a_f/a_i)^\aa$. Further, demanding the continuity of $f$ at the end of inflation, we have,
\begin{eqnarray}
f&=&f_1=\left[\frac{a}{a_i}\right]^\aa~~\text{for}~~a \le a_f \\
&=&f_2=\left[\frac{a_f}{a_i}\right]^\aa\left[\frac{a}{a_f}\right]^{{-\beta}}~~\text{for}~~a_f \le a \le a_r
\end{eqnarray}

In order to calculate the evolution of the vector potential after inflation, we solve Eq.(\ref{scriptAeq}). Since we are interested in the super-horizon scales, we express the solution (denoted by $\bar{A_2}$) as,
\begin{align}
 \bar{A_2}=&d_1+d_2 \int_{\eta_f}^{\eta} \frac{1}{f_2^2} d\eta. \label{eqa2}
\end{align}
Here $\eta_f$ is the conformal time at the end of inflation. Expressing the solution in terms of $a$, we make use of the dependence of $a$ on $\eta$ during matter dominance, $a=(H_f^2 a^{3}_{f}/4)( \eta + 3/(a_f H_{f}))^2 $. We arrive at this expression by ensuring the continuity of $a$ and $a'$ at the end of inflation. \\
The solution becomes,
\begin{align}
 \bar{A_2}=&d_1+d_2 \int_{a_f}^{a} \frac{1}{a_f^{2}H_{f}\sqrt{a/a_f}f_2^2} da. \label{eqa3}
\end{align}
As $\mathcal A = f\bar{A}$ determines the growth or decay of the
magnetic and electric fields, the constant solution
$\bar{A}_2 =d_1$ is a decaying mode when $f$ decreases. 
Thus we need to have a non-zero $d_2$ to get growing modes
during the epoch after inflation when $f$ decreases back to unity.

\begin{figure}[!]
 \epsfig{figure=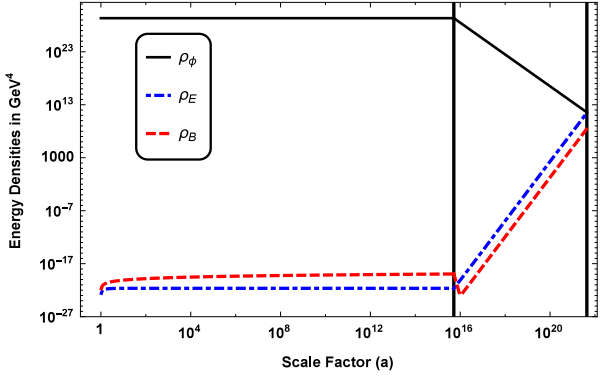,height=5cm,width=8cm,angle=0}
  \caption{\textit{In this figure evolution of $\rho_{\phi}$ , $\rho_E$ and $\rho_B$ with scale factor has been shown. We have taken $\aa=2$ and fixed $T_r$ at 100 GeV. Two vertical bold black lines represent the values of scale factor at the end of inflation $a_f$ and at reheating $a_r$ respectively. This figure shows that the energy of EM field does not overshoot the energy of the scalar field $\phi$ which decides the background geometry.}}
  \label{final plot}
\end{figure}

To find the constants $d_1$ and $d_2$, 
we demand that the value of $\bar{A}$ as well as its first derivative (with respect to conformal time) be continuous at the end of inflation. Using $ \eqref{A} $, the expression for $\bar{A}$ can be obtained during inflation ($ \bar{A}_{1} $) as,
\begin{align}
\bar{A}_1&=\frac{\mathcal A_1}{f_1} = c_1 \frac{ 2^{\frac{1}{2} + \aa } k^{-\aa} H_{f}^{\aa}}{ \Gamma (\frac{1}{2} - \aa)} \Big[1-\frac{\left(\frac{k}{a H_f}\right)^2}{4(\frac{1}{2} - \aa)}\nonumber \\ 
&+\frac{\left(\frac{k}{a H_f}\right)^4}{32(\frac{1}{2} - \aa )(\frac{3}{2} - \aa)}\Big]
+c_2 \frac{ k^{-\aa} H_f^{\aa}\left(\frac{k}{a H_f}\right)^{2 \aa + 1} }{2^{\aa + \frac{1}{2}} \Gamma(-\frac{3}{2} + \aa )}.
\end{align}
Here we have included the higher order terms as well, that are obtained by expanding the Bessel functions in the super-horizon limit. We can see from the expression above, that the first term in the $c_{1} $ branch is time independent. If we do not consider the higher order terms in the $c_{1}$ branch, it would not contribute to the coefficient $d_{2}$ during derivative matching. 
In fact,
the contribution of these higher order terms have more weightage than the $c_{2}$ branch.  Matching the expression above and its derivative to the expression of $ \bar{A}_{2} $ given in  Eq.\eqref{eqa3} and its derivative we get the coefficients $ d_{1} $ and $ d_{2} $ as,
\begin{align}
 d_1=&c_1 \frac{2^{\frac{1}{2}+\aa}\left(\frac{k}{H_f}\right)^{-\aa}}{\Gamma(\frac{1}{2}-\aa)} \left(1+\frac{\left(\frac{k}{a_f H_f}\right)^2}{2 (2 \aa -1)}+\frac{\left(\frac{k}{a_f H_f}\right)^4}{8 (2 \aa -1) (2 \aa -3)}\right)\nonumber\\
&+c_2 \frac{2^{-\frac{1}{2}-\aa}\left(\frac{k}{H_f}\right)^{-\aa}}{\Gamma(\frac{3}{2}+\aa)} \left(\frac{k}{a_f H_f}\right)^{2 \aa +1}\nonumber\\
d_2=&\Bigg[ \frac{ 2^{\frac{1}{2}+\aa}\left(\frac{k}{H_f}\right)^{-\aa}}{\Gamma(\frac{1}{2}-\aa)} c_1 \left(\frac{-k\left(\frac{k}{a_f H_f}\right)}{2\aa-1}+\frac{-k \left(\frac{k}{a_f H_f}\right)^3}{2 (2\aa-1)(2\aa-3)}\right)\nonumber\\
&+c_2\frac{2^{-\frac{1}{2}-\aa}(2\aa+1)(-k)\left(\frac{k}{H_f}\right)^{-\aa}}{\Gamma(\frac{3}{2}+\aa)} \left(\frac{k}{a_f H_f}\right)^{2\aa}\Bigg]f^2_2(a_f).
\end{align}
After substituting $d_1$ and $d_2$ in Eq.\eqref{eqa3}, we use the solution of $ \bar{A}_2$ in Eq.\eqref{rhob} and Eq.\eqref{rhoe}
for the magnetic and electric energy densities. 
Further retaining only the dominant terms, the post inflationary magnetic and electric energy density spectra at reheating, reduces to,
\begin{align}
\frac{d \rho_{B}(k,\eta)}{d \ln k}\Big|_R\approx&\frac{2^{2\aa+1}}{32 \pi \cos^{2}(\pi \aa)} \frac{1}{a_r^4 \Gamma^2(\frac{1}{2}-\aa)} \\  & \times \Bigg[\frac{ k^{-2\aa+8}H_f^{2\aa-4}}{(\frac{1}{2}-\aa)^2 a_f^4(2\beta+\frac{1}{2})^2}\Bigg]\nonumber
\left(\frac{a}{a_f}\right)^{4\beta+1} \label{rhobnew}\\
 \frac{d \rho_E(k,\eta)}{d \ln k}\Big|_R\approx&\frac{2^{2\aa+1}}{32 \pi \cos^{2}(\pi \aa)} \frac{1}{a_r^4 \Gamma^2(\frac{1}{2}-\aa)} \\ \nonumber & \times \Bigg[\frac{ k^{-2\aa+6}H_f^{2\aa-2} f_2^4(a_f)}{a_f^2(\frac{1}{2}-\aa)^2}\Bigg].
\end{align}
If we consider $\aa = 2$ case which implies a scale invariant magnetic and electric energy density spectrum during inflation, we get a blue magnetic spectrum,
with $d \rho_{B}(k,\eta)/(d \ln k) \propto k^4$,
after inflation. Both the magnetic and electric energy density increase after inflation in this case with the electric energy density dominating over the magnetic energy density. This is evident from Fig.(\ref{final plot}),
where we show the evolution of $\rho_B$ (red dashed line), $\rho_E$ 
(blue dashed-dotted line) and $\rho_\phi$ (black solid line).

It is more convenient to express the ratios of scale factors in terms of the number of e-foldings. Let $N$ denote the number of e-foldings from the beginning to the end of inflation and $N_r$ denote the number of e-foldings from the end of inflation to the reheating era. Thus the ratios of scale factors become,
\begin{equation}
\frac{a_f}{a_i}=e^{N}~~~\text{and}~~~\frac{a_r}{a_f}=e^{N_r}. \label{N}
\end{equation}
By demanding $f(a_r)=1$, we can express $\beta=\alpha N/N_r$.
We need to ensure that the total energy density in the electric and magnetic fields  does not exceed the total energy density in the inflaton field. During inflation, this condition is always satisfied for 
the particular values of $\aa$ that we consider,
for which back reaction problem does not exist. 
The same is shown in the Fig.(\ref{final plot}) for $\aa=2$, the scale invariant case.
The total electromagnetic energy density at the end of reheating is,
\begin{align}
 \rho_E+\rho_B \Big|_R&=\int_{a_i H_f}^{k_r} \frac{d \rho_E(k,\eta)}{d \ln k} d \ln k+\int_{a_i H_f}^{k_r}
\frac{d\rho_B(k,\eta)}{d\ln k} d \ln k \nonumber\\
 &\approx (C+D) H_f^4 e^{\aa (2N+N_r)-7N_r},    \label{46}
\end{align}
where $k_{i}= a_{i}H_{f}$ to $k_{f} = a_{f}H_{f}$ represents the modes which leave the horizon during inflation. Post-inflation, in the matter dominated era, a range of modes re-enter the horizon. The mode which enters the horizon at reheating is $ k_r= a_r H_r $. Therefore, at reheating the super-horizon modes,
which have been amplified have wave numbers between
$ k_{i}$ to $k_{r} $. In the above expression we have substituted  $k_r=a_r H_r=a_f H_f e^{-N_r/2}$. The value of $H_{r}$ is obtained by evolving the Hubble parameter during matter dominance. \\
The coefficients $C$ and $D$ are respectively,
\begin{align}
C=&\frac{2^{2\aa+1}}{32 \pi (\frac{1}{2}-\aa)^2 (-2\aa+6) \Gamma^2(\frac{1}{2}-\aa) \cos^2(\aa \pi)}\nonumber\\
D=&\frac{2^{2\aa+1}}{32 \pi (\frac{1}{2}-\aa)^2 (2\beta+\frac{1}{2})^2 (-2\aa+8)\Gamma^2(\frac{1}{2}-\aa) \cos^2(\aa \pi)}\nonumber  
\end{align}
The background energy density at the end of reheating is given by,
\begin{align}
\rho_\phi|_r= g_r \frac{\pi^2}{30}T_r^4.
\end{align}
Here $g_r $ represents the relativistic degrees of freedom at reheating and $T_r$ is the temperature at reheating. Hence by imposing the condition $ \rho_{B} + \rho_{E} < \rho_{\phi} $ and substituting the value of $ \rho_{B} + \rho_{E}$ from Eq.\eqref{46},
we get the following constraint,
\begin{align}
2\aa(N+N_r)-(7+\aa)N_r<\ln\left(\frac{\pi^2 g_r}{30(C+D)}\right)-4\ln\frac{H_f}{T_r}.\label{bound1}
\end{align}
In the above expression, $N$ and $N_r$ are not independent of $H_f$ and $T_r$. They are related by the fact that the present observable universe has to be inside the horizon at the beginning of inflation to explain the isotropy of Cosmic Microwave Background Radiation (CMB). This condition implies,\\
\begin{align}
\quad (a_{0}H_{0})^{-1} &< (a_{i}H_{f})^{-1} \nonumber \\
 \frac{1}{H_0}\frac{a_r a_f a_i}{a_0 a_r a_f}&<\frac{1}{H_f}.
\end{align}
Here, $ a_{0} $ and $H_{0} $ represent the scale factor and Hubble parameter today, respectively. By relating the ratios of scale factors to the number of e-folds, we get the following constraint.
\begin{align}
 N+N_r&>66.9-\ln\left(\frac{T_r}{H_f}\right)-\frac{1}{3}\ln\frac{g_r}{g_0}.\label{nnr}
\end{align}
To get the above expression, we have assumed radiation dominated era from reheating till today. The scale factors $a_{0}$ and $a_{r} $ are related by,
\begin{equation}
\frac{a_0}{a_r}=\left(\frac{g_r}{g_0}\right)^{\frac{1}{3}} \frac{T_r}{T_0}.
\end{equation} 
Here $g_0$ represents relativistic degree of freedom today and $T_0$ is the CMB temperature today.

$N_r$ can be written in terms of $H_f$ and $T_r$ using the inflaton energy density at the end of inflation($\rho_{inf}$) and reheating (${\rho_{\phi}|_r}$),

\begin{align}
N_r=\frac{1}{3} \ln \frac{\rho_{inf}}{\rho_{\phi}|_r}=\frac{1}{3} \ln\left[\frac{90 H^2_{f}}{8 \pi G \pi^2 g_r T_r^4}\right]\label{nr}
\end{align}
Substituting Eq.$\eqref{nnr}$ and Eq.$\eqref{nr}$ in  Eq.\eqref{bound1} and writing $N_r$ in terms of $H_f$ and $T_r$, we get
the constraint
\begin{align}
&\ln\Bigg[\frac{C+D}{g_r}\left(\frac{g_0}{g_r}\right)^{\frac{2\aa}{3}}\left(\frac{g_r \pi^2}{30}\right)^{\frac{7+\aa}{3}}\Bigg]+134\aa+(2\aa+4)\ln\frac{H_f}{T_r}\nonumber\\
&-\frac{4(7+\aa)}{3}\ln\left(\sqrt[4]{\frac{3 H_f^2}{8\pi G}}\frac{1}{T_r}\right)<0.  \label{bound}
\end{align}
If reheating temperature and the scale of inflation satisfy the above bound, our prescribed model will be able to circumvent the strong coupling problem during inflation and back reaction in the post inflationary era. 

From the inequality in Eq.(\ref{bound}), for a particular value of $\aa$, once we fix $T_r$, we can get an upper bound on $H_f$ or vice-versa. After we know the above three quantities, we can proceed to calculate $N_{r}$ from Eq.\eqref{nr}. This value of $N_r$ is also an upper bound. Together with the above information, we can estimate $N$ from the bound in Eq.\eqref{nnr}.
We can further estimate the proper coherence length at reheating using the expression below,
\begin{align}
L_c= a_r \frac{\int_{0}^{k_r} \frac{2 \pi}{k} \frac{d \rho_B(k,\eta)}{d \ln k} {d \ln k}}{\int_{0}^{k_r} \frac{d \rho_B(k,\eta)}{d \ln k} d \ln k}.
\label{Lc}
\end{align}
For this coherence length we can estimate the magnetic field strength at reheating using,
\begin{align}
B[L_c]= \sqrt{8 \pi \frac{d \rho_B(k,\eta)}{d \ln k}}\Big|_{k=\frac{2\pi a_r}{L_c}}.
\label{BLc}
\end{align}

\section{Predicted magnetic field strength and coherence length due to flux freezing evolution}\label{result}

After reheating, the universe is composed of conducting relativistic plasma.
The electric fields produced in the previous epochs get shorted out.
As far as the magnetic fields are concerned,
 several processes affect their evolution.
The simplest of them is the expansion of the universe where $B \propto 1/a^{2}$. However, on sub-Hubble scales, nonlinear processes in plasma also play an important role in their evolution. In this section we just consider the former case and the effect of nonlinear processes will be considered in the next section.

We consider reheating at different temperatures ($T_r$). The lowest reheating temperature we consider is 5 MeV as reheating below this energy scale 
is ruled out by Big Bang Nucleosynthesis 
constraints \cite{bbn}. 
We 
carry out the 
analysis as mentioned in  
section \ref{ascp} and estimate the coherence length ($L_c$) and magnetic field strength ($B[L_c]$) at reheating. To estimate the present day value of the magnetic field strength and its corresponding coherence length, we consider their evolution to be given by,
\begin{align}
L_{c0}=& L_c\left(\frac{a_0}{a_r}\right), \nonumber \\ 
B_0[L_{c0}]=& B[L_c]\left(\frac{a_0}{a_r}\right)^{-2},   
\label{BL_linear}
\end{align}
where $L_c$ and $B(L_c)$ 
are as in Eq.(\ref{Lc}) and Eq.(\ref{BLc}) respectively.
Our analysis is repeated for a number of possible reheating temperatures,
starting from $T_r=5$ MeV to $T_r= 1$ TeV. Of particular interest
are the epochs of reheating at QCD phase transition ($T_r=150$MeV)
and electroweak phase transition ($T_r=100 $GeV).
These cases have been considered for $\aa=2$ which gives a scale invariant magnetic field spectrum during inflation. Other values of $\aa$ can also be considered if we allow a departure from scale invariance during inflation while at the same time ensuring that the backreaction problem does not arise. As an example we have considered $\aa=3$. The latter case gives us a 
$d\rho_B/d(\ln(k)) \propto k^2$ magnetic spectrum 
instead of a  $k^4$ 
spectrum 
for super-horizon modes in the post inflationary era. We note that for GUT 
scale inflation ($10^{14}$ GeV), 
satisfying all the constraints requires a
reheating temperature 
$\approx 10^{-8}$ GeV, which falls below the temperature prescribed by the BBN bound. 

The results are 
given
in Table \ref{table1}. As evident from the table, for a reheating temperature at 100 GeV, the magnetic field strength is $5.6 \times 10^{-7}$ G at a coherence length of $ 8.8 \times 10^{-10} $ Mpc. The magnetic field strength and the coherence length, both 
increase,
as reheating temperature 
decreases to 5 MeV from 1000 GeV. 
Deviating from scale invariance during inflation, for $\aa=3$, we find that at QCD phase transition (150 MeV), the magnetic field strength achieved is 
$9.2 \times 10^{-9}$ G at a coherence length $ 9.7 \times 10^{-7}$ Mpc.
The magnetic field strength is lower than that obtained for $\aa=2$ although the coherence length obtained is 
larger.

Note that in all our models, the magnetic spectrum is blue and in addition,
the coherence scale also becomes smaller than the Hubble radius.
Therefore it becomes necessary to consider the nonlinear processing 
and damping of the magnetic field, over and above its flux freezing evolution.
This is taken up in the next section.
\begin{widetext}
\begin{center}
\begin{table}[t]
\caption{Present day magnetic field strength and coherence length in different models (for different values of $\aa$ and $T_r$), assuming only flux freezing evolution.}
\begin{tabular}{|p{1.6 cm}|l| p{2 cm}|l|l|l| p{2.2 cm}|p{2 cm}|p{2cm}|p{2cm}|l|l|}
\hline
Scale of inflation (in GeV) &$H_f$ (in GeV) &Reheating Temperature $T_r$&${\aa}$&$ N$ &~~~ $N_r $~~~&Coherence length $L_{c0}$ (in Mpc)&Magnetic field strength $B_0[L_{c0}]$(in G)\\
\hline
$1.99\times10^{11}$&$9.36\times10^{3}$&$5$~~~~~MeV&2&39.73&41.33&$2.59\times10^{-5}$&$2.04\times10^{-6}$\\
\hline
$4.94\times10^{9}$&$5.80$&$150$~~MeV&2&38.41&31.28&$6.45\times10^{-7}$&$1.28\times10^{-6}$\\
\hline
$6.77\times10^{6}$&$1.09\times10^{-5}$&$100$~~GeV&2&36.18&13.64&$8.84\times10^{-10}$&$5.59\times10^{-7}$\\
\hline
$6.77\times10^{5}$&$1.09\times10^{-7}$&$1000$~GeV&$2$&$35.42$&$7.50$&$8.84\times 10^{-11}$&$3.25\times10^{-7}$\\
\hline
$6.33\times10^{2}$&$9.51\times10^{-14}$&$5$~~~~~MeV&3&26.70&15.24&$3.89\times10^{-5}$&$2.63\times10^{-8}$\\
\hline
$1.03\times10^{2}$&$2.52\times10^{-15}$&$150$~~MeV&3&26.61&7.70&$9.69\times10^{-7}$&$9.20\times10^{-9}$\\
\hline
\end{tabular}
\label{table1}
\end{table}
\end{center}
\end{widetext}

\section{Nonlinear evolution of magnetic field}\label{nl}

The nonlinear evolution of tangled small scale magnetic fields
has been extensively discussed by Banerjee and Jedamzik 
\cite{jedamzik} (see also \cite{kandu2016}). 
We first summarize their arguments and then 
apply their results to our magnetogenesis scenarios. 

Nonlinear processing becomes important 
when the Alfv\'en crossing time ($\eta_{NL}=(kV_A(k))^{-1}$) of a mode becomes smaller than the Hubble time ($H^{-1}$).
Here $V_A(k) \equiv \sqrt{(d \rho_B/d \ln k)/(\rho+p)}$ is the Alfv\'en velocity at $k$. 
The energy density and pressure of the relativistic species are denoted by $\rho~ \text{and}~ p$.
The Lorentz force due to the field can then drive fluid motions to the
Alfv\'en velocity within an expansion time, provided 
the viscosity of the fluid is small enough. This is indeed the case
in most epochs. 
The fluid Reynolds number $R_e$ is then typically large leading to cascade of energy to smaller and smaller scales or in other words a state of MHD turbulence.
When such a state is achieved, energy at 
the coherence scale
is transferred to small scales down to the dissipation scale. Since the kinetic energy comes from initial magnetic energy and there is no other energy source, magnetic energy also decays. For a blue spectrum, 
since the Alfv\'en crossing time increases with scale $L$, the energy at the next largest scale starts dominating the spectra
and this scale now becomes the new coherence length. 
The detailed evolution needs to be studied numerically including the
effect of intervening epochs when viscosity is important.
The net result of the nonlinear processing during the radiation dominated
epochs can be summarized by the following evolution 
equations for the proper magnetic field $B^{NL}[L_{c}]$ and proper coherence
length $L_{c}^{NL}$ 
\cite{jedamzik}, 
\begin{align}
B_0^{NL}[L^{NL}_{c0}] = B_0[L_{c0}]\left(\frac{a_m}{a_r}\right)^{-p} \text{,} ~~L_{c0}^{NL} = L_{c0} \left(\frac{a_m}{a_r}\right)^{q},
\end{align}
where $a_m$ is the scale factor at radiation-matter equality, $p \equiv (n+3)/(n+5)$ and $q \equiv 2/(n+5)$. Here, $n$ is defined in such a way that $(d \rho_B/d \ln k)\propto k^{n+3}$. 
Here we have also used the fact that the expansion factor during
radiation domination varies as $a(\eta) \propto \eta$.
We consider the form of evolution mentioned above up to matter-radiation equality after which $L^{NL}_c$ grows only logarithmically \cite{jedamzik}. We neglect the logarithmic growth in $L^{NL}_c$ and evolve the two quantities till today in a similar way as mentioned in Section \ref{result}. 

The results of this calculation are shown in Table \ref{table2}. 
Compared to the previous case where nonlinear effects are not considered, we see that the coherence length is larger and the magnetic field strength is lower for the same reheating temperature. 
For reheating at QCD phase transition (150 MeV), and taking $\aa=2$ (or $n=1$),
 the magnetic field strength comes out to be $ 1.4 \times 10^{-12} G $ 
at the corresponding coherence length of $ 6.1 \times 10^{-4} $ Mpc. 
This value of coherence length 
increases for $ \aa = 3$ to $ 2.8 \times 10 ^{-2}$ Mpc 
while the corresponding magnetic field strength 
decreases
to $ 3.2 \times 10^{-13} $ G.  

\begin{widetext}
 \begin{center}
\begin{table}[h!]
\caption{Present day magnetic field strength and coherence length in different models (for different values of $\aa$ and $T_r$), after taking nonlinear effects into account.}
\begin{tabular}{|p{2 cm}|p{2 cm}|l|p{2 cm}|p{2 cm}|p{2 cm}|p{2 cm}|}
\hline
Scale of inflation (in GeV)
&Reheating Temperature $T_r$&${\aa}$&Coherence length~$L^{NL}_{c0}$ (in Mpc)&Magnetic field strength $B_0^{NL}[L^{NL}_{c0}]$(in G)&Coherence length~$L^{S}_{c0}$ (in Mpc)&Magnetic field strength $B_0^{S}[L^S_{c0}]$(in G)\\
\hline
$1.99\times10^{11}$
&$5$~~~~~MeV&2&$6.48\times10^{-3}$&$3.26\times10^{-11}$&0.102&$5.15\times10^{-10}$\\
\hline
$4.94\times10^{9}$&
$150$~~MeV&2&$6.09\times10^{-4}$&$1.43\times10^{-12}$&$1.09\times10^{-2}$&$4.38\times10^{-11}$\\
\hline
$6.77\times10^{6}$
&$100$~~GeV&2&$7.74\times10^{-6}$&$7.26\times10^{-15}$&$7.25\times10^{-4}$&$6.81\times10^{-13}$\\
\hline
$6.77\times10^{5}$
&$1000$~GeV&$2$&$1.67\times10^{-6}$&$9.11\times10^{-16}$&$2.29\times10^{-4}$&$1.26\times10^{-13}$\\
\hline
$6.33\times10^{2}$
&$5$~~~~~MeV&3&$0.153$&$6.65\times10^{-12}$&$0.153$&$6.65\times10^{-12}$\\
\hline
$1.03\times10^{2}$
&$150$~~MeV&3&$2.81\times10^{-2}$&$3.17\times
10^{-13}$&$2.81\times10^{-2}$&$3.17\times
10^{-13}$\\
\hline
\end{tabular}
\label{table2}
\end{table}
\end{center}
\end{widetext}

Numerical simulations 
by Brandenburg et. al. \cite{axel} show a slower decay of non-helical magnetic fields. This slower decay of the field is also accompanied by an apparent inverse transfer of the magnetic energy to larger scales
\cite{axel,axel2017, zrake},
 which is usually thought to occur only for helical fields. 
It would be of interest to examine the consequence of this non-helical
inverse transfer to the predicted field strengths and coherence scales. 
The simulations of \cite{axel}, start from a blue spectra with
$n=2$, and show that the magnetic field energy 
decays as $B_0^{S}[L^S_{c0}] = B_0[L_{c0}]\left(a_m /a_r\right)^{-0.5}$ 
and coherence length increases as 
$L_{c0}^{S} = L_{c0} \left(a_m/a_r\right)^{0.5}$. 
These simulations are motivated by studying the decay of causally generated
fields which typically are expected to have such a spectrum on infra-red
scales. Our inflation generated fields, although having blue
spectra, have more power on large scales, with $n=1$ for $\aa=2$ case
and $n=-1$ for the case when $\aa=3$, and thus the magnetic energy
could perhaps decay even more slowly. Nevertheless, to 
get an idea of what such a non-helical
inverse transfer would imply for the field strengths and coherence scales,
we adopt simply the scalings found by Brandenburg et al \cite{axel}.
The results are shown in
Table \ref{table2} 
for different reheating temperatures. The analysis is same as mentioned in Section \ref{result}. We note that 
both
the coherence length 
and magnetic field strength are
larger than the one estimated using only the standard nonlinear evolution. 

For example, at a reheating temperature around QCD phase transition (150 MeV), 
the coherence length 
increases from $6.1 \times 10^{-4}$ Mpc to $1.1 \times 10^{-2}$.  
Further the magnetic field strength 
increases to $ 4.4 \times 10^{-11} $ G  from $1.4 \times 
10^{-12} $ G. This is the case for $ \aa = 2$ ($n=1$). 
For the case of $ \aa = 3$ ($n=-1$) the evolution relations of the magnetic field strength and the corresponding coherence length are equivalent to those obtained 
in \cite{axel}. 
Hence, the values do not change after incorporating inverse transfer. 
We also note that, though
the magnetic field strength 
is larger 
after considering inverse transfer, it still remains far lower than the strength obtained 
when nonlinear effects are not taken into account. 
The corresponding coherence length however is 
considerably 
enhanced. 
We recall that 
for reheating at 150 MeV, the magnetic field strength 
without accounting for nonlinear decay, 
is $1.3 \times 10^{-6}$ G at a coherence length of $ 6.5 \times 10^{-7} $ Mpc. Thus the field strength after taking account of the nonlinear decay with
inverse transfer is much smaller ($ 4.4 \times 10^{-11} $ G), 
and the coherence scale is much larger ($1.1 \times 10^{-2}$ Mpc).
\section{Constraints from $\gamma$-ray obervations}\label{discussion}

We ask if the generated fields can explain the constraints from the gamma ray bounds. The $\gamma$-ray observations of TeV blazars 
suggest a lower limit on the strength of the intergalactic magnetic fields of the order of $10^{-15}$ G at a 
comoving
coherence length of 0.1 Mpc \cite{neronov}. 
This 
bound was obtained from the non-detection of secondary gamma ray emission by the Fermi telescope. The above mentioned lower limit was obtained for the case 
$L_{C} \gg L_{IC}$, where $L_{C}$ is the 
proper
coherence length of the magnetic field and $L_{IC}$ is the mean free path of the charged particles that undergo inverse compton scattering. 
For $L_{C} \ll L_{IC}$, 
the lower limit on magnetic field strength increases with coherence length as $ (L_c)^{-1/2}$. 

\begin{figure}[h!]
 \epsfig{figure=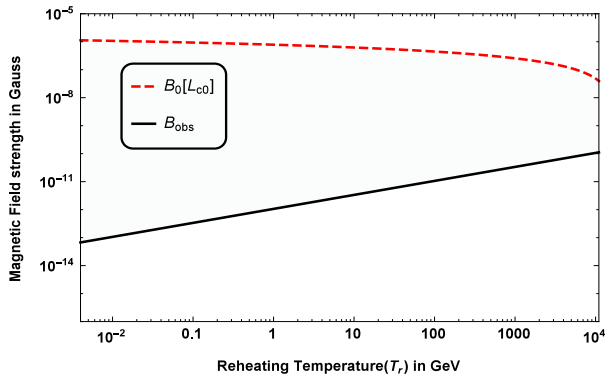,height=5cm,width=8cm,angle=0}
  \caption{ \textit{This figure is for ${\aa}=2$. The red dashed curve represents the maximum magnetic field strength $B_0[L_{c0}] $ that can be generated in our model without taking nonlinear evolution. The black curve shows the lower bound on the observed magnetic field strength constrained by the gamma ray observations ($ \ge 10^{-15} ~\text{at}~ 0.1 Mpc $) for different values of $T_r$. They are estimated at the coherence length of the generated magnetic field. The shaded region represents all the observationally allowed magnetic field strengths permitted by our model.}}
 \label{fig2}
 \end{figure}

\begin{figure*}
 \epsfig{figure=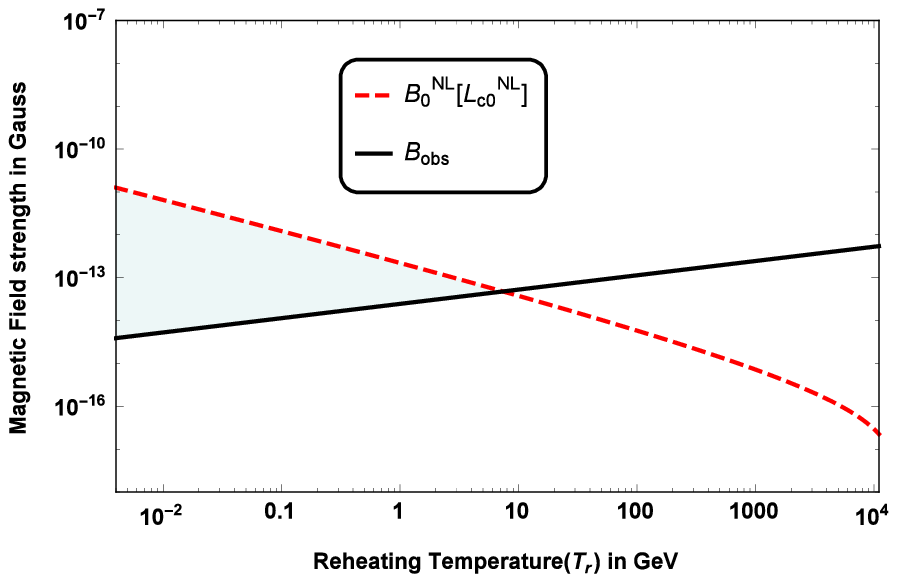,height=5cm,width=8cm,angle=0}
\epsfig{figure=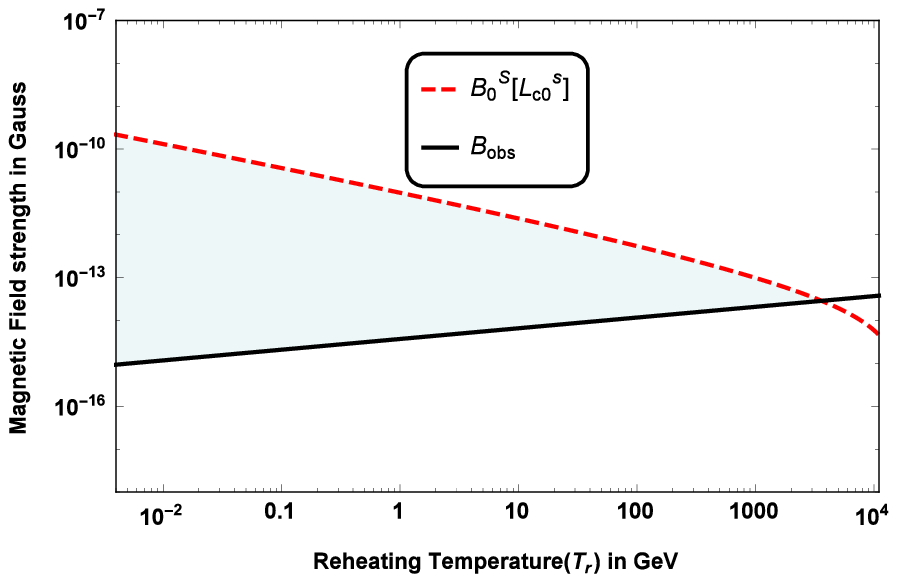,height=5cm,width=8cm,angle=0}
  \caption{{\textit{These figures are for ${\aa}=2$. The black curve in both the figures represents the lower bound on observed magnetic field strength constrained by the gamma ray observations ($ \ge 10^{-15} $ G at  $0.1 Mpc $) for different values of $T_r$. They are estimated at the coherence length of the generated magnetic field. The red dashed curve in the left panel represents the maximum magnetic field strength ($B_0^{NL}[L^{NL}_{c0}]$) that can be generated in our model by taking nonlinear evolution. While the red dashed curve in the right panel represents the maximum magnetic field strength ($B_0^{S}[L^S_{c0}]$) that can be generated by taking direct numerical simulations results for the nonlinear evolution of magnetic fields. The shaded region in both the figures represents all the observationally allowed magnetic field strengths permitted by our model.}}}
\label{fig3}
\end{figure*}


In Fig.(\ref{fig2}) and Fig.(\ref{fig3}) we compare the lower bound
on the fields obtained from the $\gamma$-ray observations (black solid
line), with the predicted field strength from our models at the
coherence scale, for different reheating temperatures (red dashed line).
The black line in the figures, which represents the gamma ray bound, was 
evaluated by scaling $10^{-15}$ G at 0.1 Mpc to the coherence length 
obtained for a particular reheating temperature. The scaling relation 
as mentioned above goes as $ (L_c)^{-1/2}$. 
Fig.(\ref{fig2}) is for the case when only the flux freezing evolution
is taken into account (without nonlinear processing).
 The left panel
of Fig.(\ref{fig3}) is for the case when the standard nonlinear evolution
is taken into account, while the right panel assumes the scalings
implied by a possible nonlinear inverse transfer. All the figures assume
$\aa=2$, where the spectra are scale invariant during the inflationary
era and transit to a blue spectrum with $n=1$ by reheating.

We can see from Fig.(\ref{fig2}), that for the range of reheating temperatures we have considered, the magnetic field strength obtained for a particular coherence length lies well above the lower limit prescribed by the gamma ray observations. For $\aa =2 $ case, the reheating temperatures can range from a minimum value of 
5 MeV to a maximum limit of $ 1.7 \times 10^{4}$ GeV. 
For $\aa = 3$ case, this limit decreases to $6.7$ GeV. This is because the magnetic energy density diverges during inflation and to prevent it from exceeding the inflaton energy density requires a low-scale inflation. Further to satisfy the constraint in Eq.\eqref{bound}, reheating temperature also has to be low.

Taking into account nonlinear evolution tightens these constraints for both $\aa=2$ and $\aa =3$. The change in the constraint for $\aa=2$ case, can be seen in Fig.(\ref{fig3}). 
The shaded region shown in the figure corresponds to the values of magnetic fields allowed by the gamma ray bounds.
If non-helical decaying MHD turbulence
does not have an inverse transfer (left panel of Fig.(\ref{fig3})),
the reheating temperature has to be below $\approx 7$ GeV for
the $\gamma$-ray bound to be satisfied.  
On the other hand, if one takes into account the 
inverse transfer as discussed in Section \ref{nl}, 
we saw that larger magnetic field strengths are possible.
Then from the right panel of Fig.(\ref{fig3}), we see that 
the limits on the reheating temperatures get relaxed.
In this case, the reheating temperatures allowed by the gamma ray 
bound increases to $\approx 4 \times 10^{3} $ GeV. 
For $\aa = 3$ case, the maximum reheating temperature is not drastically affected by considering nonlinear evolution. It decreases from $6.7$ GeV to $4$ GeV. This limit is also not affected by including the inverse transfer phenomena and $T_{r}$ remains at $4$ GeV. In case the lower limit on the magnetic field strength from $\gamma$-ray observations is decreased to $10^{-17}$ G at $1$ Mpc as discussed in \cite{2011A&A...529A.144T}, our constraint on the reheating temperature will be relaxed further.

Thus we see that to resolve
the strong coupling problem 
(by demanding $f(\phi)$ to decay to its pre-inflationary value) 
and also to avoid back reaction in the post inflationary era, 
requires both a low scale inflation and a low scale of reheating. 
These models however lead to magnetic field strengths and
coherence scales which are consistent with the $\gamma$-ray lower limits,
for reheating temperatures up to about few thousand GeV.

We note that few earlier studies have also considered 
low-scale inflation in the context of 
inflationary
magnetogenesis. Ferreira et. al. \cite{rajeev} have discussed a scenario 
wherein they have considered the back reacting branch of $f(\phi)$ during a period of low-scale inflation. Although the model successfully satisfies the gamma ray bound and avoids both strong coupling and back reaction, it violates the Schwinger effect constraint as discussed by Kobayashi and Afshordi \cite{kobayashi:2014}. Our model on the other hand  does not run into Schwinger effect inconsistencies as discussed in Section \ref{se}.


\section{Discussion and Conclusions}\label{conclusion}

We have studied here the generation of magnetic fields during the inflationary era.
As the standard Maxwell action is conformally invariant, 
electromagnetic (EM)
field fluctuations decay with expansion as 
$1/a^{2}$. 
Thus to generate fields of significant 
present day strengths, breaking of conformal invariance is imperative. 
One of the ways 
this can be 
done is by coupling the EM action to a function of the inflaton ($f^2FF$ \cite{ratra}). Although 
such a model can lead to 
the generation of magnetic fields with 
present day strengths of interest, 
it suffers from 
several potential problems.
These have been referred to as the 
strong coupling problem, back reaction problem and the 
Schwinger effect constraint. 

In Section \ref{emdi}, we have shown that in the $f^2FF$ model, 
there are two possible evolution of $f(\phi)$ to generate scale 
invariant magnetic field spectrum. In the first case, $f$ increases as $a^2$ but if one demands that 
the effective electric charge is restored to the standard value 
at the end of inflation, the model 
suffers from the strong coupling 
problem at the beginning of inflation. In the second case, $f$ decreases 
as $a^{-3}$. In this case however,  the resultant electric field spectrum diverges. Hence, the model suffers from the back reaction 
problem and Schwinger effect inconsistencies.

We have proposed a model which evades all the above mentioned problems. In our model, during inflation, $f$ increases as a power law with an exponent $\aa$. We constrain $\aa$ to be greater than 1/2 and also such that there are no back reaction effects during inflation. The coupling function ($f$) 
is assumed to begin with a standard value of unity  
and it increases to a large 
value at the end of inflation. To get back the standard coupling ($e$), 
we introduce a transition at the end of inflation in the evolution of $f$. During the second part of the evolution in the post-inflationary matter dominated era, $f$ decreases as a power law with an exponent $\beta$. By demanding that the EM energy density does not back react on the background post inflation as well, we have put a bound on the reheating temperature and the scale of inflation. We note as the reheating temperature increases, the scale of inflation decreases according to the constraint in Eq.(\ref{bound}). 
Hence, the maximum reheating temperature possible is when it becomes equivalent to the scale of inflation. 
For the scale invariant magnetic spectrum during inflation i.e $\aa =2$, the upper bound on the reheating temperature obtained is $\approx 10^4$ GeV.

For each reheating temperature and the scale of inflation, we have estimated the present day magnetic field strength and the corresponding coherence length. We have considered 
different cases in the evolution of the magnetic field after reheating. 
To begin with, 
we do not consider any nonlinear effects arising due to interaction of the magnetic modes with the plasma. For this case, we obtain fields of the order 
$1.3 \times 10^{-6}$ G and the coherence length scales of the order 
$6.5 \times 10^{-7}$ Mpc for a reheating temperature at the QCD epoch ($150$ MeV) ( Refer to Table \ref{table1}). From Fig. \ref{fig2}, we 
see that for all reheating temperatures below about $10^4$ GeV, 
the magnetic field strengths and coherence scales are large enough to 
satisfy the gamma ray bound.

Taking nonlinear evolution 
and the resulting turbulent decay into 
account, 
we find that 
the $\gamma$-ray observations lead to
an upper bound on 
the reheating temperature 
of about 7 GeV 
(the left panel of Fig (\ref{fig3})). A model which does satisfy the $\gamma$-ray constraint, 
is when reheating occurs at the QCD epoch.
The coherence length is enhanced to $6.1 \times 10^{-4}$ Mpc 
and the magnetic field strength is decreased to about $1.4 \times 10^{-12}$ G.

However there is also the phenomena of inverse transfer \cite{axel} 
(predicted by numerical simulations) of non-helical MHD turbulence 
which needs to be taken into account. When this 
is also incorporated in our calculations, we see an improvement in the constraint on reheating temperatures and strength of the magnetic fields. The upper bound on reheating temperature increases to $\approx 4 \times 10^3$ GeV. For a reheating temperature at 100 GeV, the coherence length is enhanced to 
$7.3 \times 10^{-4}$ Mpc from 
$8.8 \times 10^{-10}$ Mpc (which is obtained if we assume only
pure flux freezing evolution). 
On the other hand the magnetic field strength at the above mentioned 
coherence scales decreases to $6.8\times 10^{-13}$G from 
$5.6 \times 10^{-7}$G.

One possible scenario whereby the coupling function $f$ transits 
from a growing function to one that decays back to unity can perhaps be
realized 
in models of hybrid inflation \cite{linde}. In hybrid inflation, two interacting scalar fields, $\phi$ and $\sigma$, are employed. During inflation, the inflaton field ($\phi$) slow rolls and the other field ($\sigma$) is static. To end inflation, $\phi$ triggers $\sigma$ to rapidly roll down the potential. 
The transition in $f$ can be explained by making $f$ a function of both these scalar fields. The function $f$ could increase as 
$\phi$ slow rolls during inflation 
but at the end of inflation, it shifts to a function of $\sigma$. 
It is brought down to its initial value as $\sigma$ cascades down.
More work is needed to explore this idea further.

In our analysis, we have only looked at the generation of non-helical magnetic fields. Helical magnetic fields on the other hand will provide a much bigger advantage since inverse cascade 
leads to an even milder decrease in 
the strength of the magnetic fields while increasing the coherence lengths of the fields. We intend to look at this case in the future. There is also the question of gravitational waves production from anisotropic stresses of these generated magnetic fields as discussed by Caprini and Durrer \cite{caprini}.  
It would be of interest to study gravitational wave generation 
in our model and check whether 
they could be detected in future 
generation of detectors and thus
probe magnetogenesis. 


\section*{Acknowledgments}
RS, SJ and TRS acknowledge the facilities at IRC, University of Delhi as well as the hospitality and resources provided by IUCAA, Pune where part of this work was carried out. The research of RS is supported by SRF from CSIR, India under grant 09/045(1343)/2014-EMR-I. The research of SJ is supported by UGC Non-NET fellowship, India. TRS acknowledges the Project grant from SERB EMR/2016/002286.

\newpage

\bibliographystyle{apsrev4-1}
\bibliography{references}

\end{document}